\documentclass[final,5p,times,twocolumn,authoryear]{elsarticle}

\usepackage{amsmath}
\usepackage{amsfonts}
\usepackage{amssymb}
\usepackage[titletoc,title]{appendix}
\usepackage{booktabs}
\usepackage[mmddyyyy]{datetime}
\usepackage{dutchcal}
\usepackage{enumitem}
\usepackage{graphicx}
\usepackage{hyperref}
\usepackage[utf8]{inputenc}
\usepackage{lineno}
\usepackage[percent]{overpic}
\usepackage{pgfplots}
\usepackage{relsize}
\usepackage{subfigure}
\usepackage{siunitx}
\usepackage{tikz}

\modulolinenumbers[5]
\numberwithin{equation}{section}

\usetikzlibrary{calc}
\usetikzlibrary{plotmarks}
\usetikzlibrary{positioning}
\usetikzlibrary{fit}
\usetikzlibrary{decorations.pathmorphing}
\usetikzlibrary{backgrounds}
\pgfplotsset{compat = newest}
\pgfplotsset{ legend style={font=\tiny} }
\definecolor{bgreen}{rgb}{0.0,0.5,0.0}
\definecolor{bblue}{rgb}{0.0,0.0,0.9}
\definecolor{bgold}{rgb}{0.7,0.5,0.0}
\definecolor{bred}{rgb}{0.9,0.0,0.0}

\begin{document}

\begin{frontmatter}



\title{Gravitational polarization of test-mass potential in the self-gravitating isothermal gases and a relation with gravitational instability}


\author{Yuta Ito}
\ead{yutaito30@gmail.com}

\begin{abstract}
The present work analyzes perturbed potentials due to test mass that is added at the center of gravity of the non-singular equilibrium isothermal self-gravitating gases. We examine gravitational polarization in the infinite isothermal sheet, cylinder, and sphere, assuming that the systems are highly collisional and reach a new state of thermal equilibrium after perturbation. Under the assumptions, the isothermal sheet and cylinder amplify gravitational fields due to test sheet and line masses  by 68 $\%$ and 53 $\%$ maximally. On the one hand, in the isothermal sphere,  gravitational fields due to test point mass are amplified oscillatorily with radius and show a repulsive effect at large radii.  Since the infinite isothermal sphere is  gravitationally unstable, we confine it in a spherical wall in contact with a thermal bath. We find that only gravitational amplification occurs if the unperturbed finite sphere is gravitationally stable. The oscillatory amplification appears if the sphere is unstable, hence the repulsive effect is unrealistic in the canonical ensemble.
\end{abstract}


\end{frontmatter}

\section{Introduction}

Due to the nature of long-range interactions caused by Newtonian pair-wise potentials of particles,  self-gravitating systems exhibit exotic statistical aspects, such as negative specific heat, violent relaxation, statistical ensemble inequality, and so on \citep{Dauxois_2002, Campa_2014}. One of the most fundamental but little-discussed collective effects is the amplification of gravitational field due to test mass that is added in self-gravitating systems.  The potential due to test mass attracts and reconfigures ambient masses, and reconfigured masses can amplify gravitational field due to test mass. This amplification is the counterpart of the Debye-shielding in plasmas, so it may be called ``gravitational polarization" \citep{Miller_1966, Gilbert_1970}. Gravitational polarization is expected to occur in every self-gravitating system.  It can help understand the effects of an emerging mass such as a growing black hole \citep{Young_1980, Murali_1998}, polarization of galactic halo \citep{Moody_1999}, and interactions via amplified gravitational fields \citep{Heyvaerts_2010, Chavanis_2012} in self-gravitating systems. However, the polarization effects were discussed only for collisionless systems, such as galaxies and dark matters, and for weakly collisional systems, such as globular clusters and nuclear star clusters. Those discussions were carried out in mathematically tractable and/or physically limited settings, otherwise using extensive numerical methods. 
  
In collisionless self-gravitating systems, relaxation time scale is much longer than the dynamical time scale \citep[e.g.,][]{Binney_2011}. The dynamics can be generally described by the Vlasov equation in conjunction with the Poisson equation. The simplest physical setting for gravitational amplification is to add point test mass in an infinite homogeneous collisionless self-gravitating system composed of particles obeying the Maxwellian distribution function (DF) \citep{Marochnik_1968, Padmanabhan_1985}. However, this setting necessarily faces the ``Jeans swindle" problem which is a controversy method but may be valid on spatial scales of over 100 Mpc with the cosmological principle \citep{Falco_2013}. The origin of this problem is that homogeneous self-gravitating systems can not be correctly described by the Vlasov-Poisson system. Indeed, actual self-gravitating systems are naturally finite and inhomogeneous. The first rigorous study on an infinite inhomogeneous system is due to \citep{Gilbert_1970}. Gilbert analyzed gravitational amplification for collisionless isochrone. The work showed that the effective mass, the partial sum of reconfigured masses due to a test-mass perturbation, is amplified maximally by a factor of 2.75. Another rigorous work is to analyze point-like perturbations in the singular isothermal stellar systems \citep{Murali_1998} and \citep{Murali_1999} for the King model based on \cite{Kalnajs_1977}'s matrix method.  \cite{Murali_1998} showed that a slowly growing black hole potential can be amplified by a factor of 2 in the singular isothermal sphere. The matrix method is powerful for linear response theory and applies to any unperturbed regular density using a combination of series expansion methods \citep[e.g.,][]{Binney_2011}. However, it can not manage nonlinear perturbations and needs tedious numerical and analytical calculations.

Gravitational amplification may affect slow relaxation process in weakly collisional systems. The systems reach Virial equilibria on dynamical time scales $t_\text{D}$ ($\sim $0.1 Myr) due to a rapid change in the self-consistent mean-field potential \citep{Spitzer_1988,Binney_2011}. Actual stellar systems are composed of finite number $N$ of stars against the assumption of the mean-field approximation, $N\to\infty$. On the order of $N t_\text{D}$ ($\sim$1 Gyr), stellar motions deviate substantially from the original smooth orbits determined by the mean-field potential and its cumulative change in velocity dispersion becomes compatible with the average speed of stellar motion. This process is considered slow relaxation and properly described by a generalized kinetic equation for star clusters \citep{Gilbert_1968}. It was derived by a small-parameter-$1/N$ expansion of $N$-body DF in the $N$-body Liouville equation. The characteristics of gravitational amplification is supposed to appear as a perturbed potential between ``dressed particles" in the correlation function of the kinetic equation, like the Yukawa potential in the correlation function of the Balescu-Lenard equation for plasmas \citep[e.g.,][]{Montgomery_1964}. However, because of its mathematical complication \citep{Heyvaerts_2010, Chavanis_2012}, there is little quantitative and qualitative analysis has been discussed on gravitational amplification. Exceptions are dynamical or wave phenomena in homogeneous stellar systems \cite{Weinberg_1993} and self-gravitating discs \citep{Fouvry_2017b} rather than a static configuration.   

A common problem in analyzing gravitational amplification in collisionless and weakly-collisional systems is that we must know the explicit analytical forms of orbital periods and isolating integrals for orbiting masses. This can be achieved only for limited configurations, such as isochrones \citep{Gilbert_1968}, harmonic oscillators \citep{Goodman_1984a}, and Keplerian potential. To handle the problem for more realistic systems, we have to employ \cite{Kalnajs_1977}'s matrix method, which costs exhaustive numerical and series-expansion calculations. To avoid these problems but still deepen our understanding of gravitational amplification, the present work employs collisional gaseous self-gravitating models, specifically the equilibrium isothermal systems. As mentioned in \citep{Murali_1998}, gaseous (fluid) systems are mathematically simpler than collisionless systems and could show the basic properties of gravitational amplification observed in collisionless system.

In collisional self-gravitating systems, direct head-on collisions between molecules are dominant over dynamical effects in the zeroth-order approximation. For example, assume a typical dense molecular cloud of dimension of 0.1 pc and density of $10^8\text{m}^{-3}$ and composed of $H_{2}$ molecules of radius of $0.1$ nm. The Knudsen number Kn is $\sim10^{-8}$. A highly collisional (Kn$\ll1$) self-gravitating system can reach a hydrostatic and isothermal equilibrium. Such a system is described by the Lane-Emden equation for the isothermal self-gravitating gaseous systems \citep[e.g.,][]{Horedt_2004}. The most often used are the isothermal sheet, cylinder, and sphere models. The isothermal sheet has been used to describe the vertical structure of disk galaxies \citep{Mo_2010}. The isothermal cylinder can model some filamentary structures, such as filaments in infrared-dark clouds \citep{Johnstone_2003} and baryon-rich cores of intergalactic filaments \citep{Harford_2011}. The isothermal sphere is useful to model the cores of dense molecular clouds and protostars \citep{Gnedin_2016}. To the best of our knowledge, gravitational amplification has not been analyzed for these isothermal systems. In the present work, we show that gravitational amplification occurs in all the isothermal  sheet, cylinder, and sphere while the isothermal sphere causes a repulsive gravitational effect. 

The present paper is planned as follows. Section \ref{sec_iso_models} explains the isothermal gaseous sheet, cylinder, and sphere.  Section \ref{Perturb_iso} explains the perturbation method to examine gravitational amplification in the isothermal models. Section \ref{sec:result} shows the numerical results, and Section \ref{sec:conclusion} is Conclusion.

\section{Self-gravitating isothermal gaseous models}\label{sec_iso_models}

The present section describes fundamental features of the self-gravitating isothermal sheet, cylinder, and sphere.  We first explain a general mathematical derivation of the isothermal models and then each model's feature briefly.

\subsection{Derivation of the isothermal gaseous models}\label{Deriv_Iso}

The isothermal gaseous models are typically described by a hydrostatic equation and equation of state for the ideal gas that are coupled with the Poisson equation. To compare the present work with collisionless systems, we start with the Boltzmann equation for DF $F(\boldsymbol{r},\boldsymbol{v})$ for masses $m$ interacting via pair-wise Newtonian potential at phase-space point $(\boldsymbol{r},\boldsymbol{v})$ \citep{Shu_1991}
\begin{equation}
	\left(\frac{\partial}{\partial t}+ \boldsymbol{v}\cdot\frac{\partial }{\partial{\boldsymbol{r}}} -\frac{1}{m}\frac{\partial \Phi}{\partial{\boldsymbol{r}}}\cdot\frac{\partial }{\partial{\boldsymbol{v}}}\right)F(\boldsymbol{r},\boldsymbol{v},t)=C_\text{Bol}(F,F),
\end{equation} 
where $C_\text{Bol}(F,F)$ is the Boltzmann collision term and $\Phi(\boldsymbol{r})$ the self-consistent mean-field potential. If head-on collisions dominate system dynamics, or Kn$\ll1$, then
\begin{equation}
C_\text{Bol}(F,F)=0.
\label{C_Bol}
\end{equation} 
A DF that satisfies equation \eqref{C_Bol} is a local Maxwellian DF described in terms of spatial density $n(\boldsymbol{r},t)$, bulk velocity $u(\boldsymbol{r},t)$, temperature $T(\boldsymbol{r},t)$, and energy $E(\boldsymbol{r},\boldsymbol{v},t)$ available to mass $m$.

The standard theory for the isothermal models assumes a hydrostatic equilibrium, which corresponds to that the time-independent collisionless Boltzmann equation is valid;
\begin{equation}
	\left( \boldsymbol{v}\cdot\frac{\partial }{\partial{\boldsymbol{r}}} -\frac{1}{m}\frac{\partial \Phi}{\partial{\boldsymbol{r}}}\cdot\frac{\partial }{\partial{\boldsymbol{v}}}\right)F(\boldsymbol{r},\boldsymbol{v})=0.\label{in_col_Bol}
\end{equation}
We introduce the following local Maxwellian DF $F_\text{o}$ that satisfies both the conditions in equations \eqref{C_Bol} and \eqref{in_col_Bol}
\begin{eqnarray}
		F_\text{o}(\boldsymbol{r},\boldsymbol{v})&=&mn_\text{c}\Omega^{D/2}\exp\left[-\frac{E_\text{o}(\boldsymbol{r},\boldsymbol{v})}{k_\text{B}T}\right],\label{DF_o}\\
	E_\text{o}(\boldsymbol{r},\boldsymbol{v})&\equiv&\frac{1}{2}m\boldsymbol{v}^{2}+\Phi_\text{o}(\boldsymbol{r})-\Phi_\text{c},\\
	\Omega&\equiv&\frac{m}{2\pi k_\text{B}T}, 
\end{eqnarray}
where $n_\text{c}$ is the central number density, $T$ the constant temperature, $k_\text{B}$ the Boltzmann constant, $\Phi_\text{o}(\boldsymbol{r})$ the mean-field potential due to DF $F_\text{o}$, and $\Phi_\text{c}$ the central potential. The exponent $D$ is the dimension of the isothermal models. For example, $D=1$ for the isothermal sheet, and $D=2$ for the isothermal cylinder.

From equation \eqref{DF_o}, the Poisson equation for the isothermal gaseous models reads
\begin{eqnarray}
	\frac{\partial}{\partial \boldsymbol{r}}\cdot\left(\frac{\partial \Phi_\text{o}}{\partial \boldsymbol{r}}\right)&=&4\pi G\int\text{d}^{3}\boldsymbol{v}F_\text{o}(\boldsymbol{r},\boldsymbol{v}),\nonumber\\
	&=&4\pi Gmn_\text{c}\exp\left[-\frac{\Phi_\text{o}(\boldsymbol{r})-\Phi_\text{c}}{k_\text{B}T}\right].
\end{eqnarray}  
Introducing dimensionless variables
\begin{eqnarray}
	      \phi&\equiv&\frac{\Phi_\text{o}(\boldsymbol{r})-\Phi_\text{c}}{k_\text{B}T},\\
	\boldsymbol{\xi}&\equiv&\frac{\boldsymbol{r}}{L_\text{c}}\equiv\left(\frac{4\pi Gm n_\text{c}}{k_\text{B}T}\right)^{1/2}\boldsymbol{r},\label{dimless_var}
\end{eqnarray}
The general form of the Lane-Emden equation for the isothermal gases is obtained
\begin{equation}
	\frac{\partial}{\partial \boldsymbol{\xi}}\cdot\left(\frac{\partial \phi}{\partial \boldsymbol{\xi}}\right)-\exp\left[-\phi\right]=0.\label{g_LE}
\end{equation}
To handle different spatial symmetries for the isothermal models, we employ the following modulus of independent variables
\begin{equation}
	\xi=|\boldsymbol{\xi}|=\begin{cases}
		   |z|,& (\text{linear symmetry})\\
		\rho,& (\text{cylindrical symmetry})\\
		   r,& (\text{spherical symmetry})
	\end{cases}
\end{equation}
where $z\in(-\infty,\infty)$, $\rho\in[0,\infty)$, and $r\in[0,\infty)$.

\subsection{Isothermal sheet model}

The isothermal sheet is a three-dimensional self-gravitating model composed of particles moving everywhere in the configuration spaces but stratified in the $z$-direction \citep{Spitzer_1942}. The sheet potential $\phi$ generally depends on the radial component as well on each sheet. The dependence, however, is not essential to discuss perturbation along the $z$-axis. Hence, we treat the isothermal sheet as  a one-dimensional model. Equation \eqref{g_LE} on one dimension along the $z$-axis reduces to the isothermal sheet model 
\begin{equation}
	\frac{\text{d}^2\phi}{\text{d}\,z^2}-\text{e}^{-\phi}=0,
	\label{LE_1D}
\end{equation} 
with boundary conditions (BCs)
\begin{equation}
	\phi(z=0)=0, \qquad \phi'(z=0)= 0. \label{BC_LE_1D}
\end{equation}
The explicit analytical solution to equation \eqref{LE_1D} is known as
\begin{eqnarray}
	\phi(z)&=&\ln\left[\cosh^{2}\left(\frac{z}{\sqrt{2}}\right)\right],\\
	   n(z)&=&\text{e}^{-\phi(z)}=\cosh^{-2}\left(\frac{z}{\sqrt{2}}\right).
\end{eqnarray}
The total number of masses is finite (per unit cross section) as follows
\begin{equation}
	N=\int_{-\infty}^{\infty}n\left(z'\right)\text{d}z'=2\sqrt{2}.
\end{equation}

\subsection{Isothermal cylinder model}
The isothermal cylinder model is also a three-dimensional self-gravitating system and was initially derived as the zeroth-order approximation of gaseous rings \citep{Ostriker_1964}.  Consider an infinitely-long cylindrical system aligned with the $z$-axis and described in terms of radius $\rho$ measured from the $z$-axis. Then, equation \eqref{g_LE} reduces to the isothermal cylinder model  
\begin{equation}
	\frac{\text{d}^2\phi}{\text{d}\,\rho^2}+\frac{\text{d}\phi}{\text{d}\,\rho}-\text{e}^{-\phi}=0,
	\label{LE_2D}
\end{equation} 
with BCs
\begin{equation}
	\phi(\rho=0)=0, \qquad \phi'(\rho=0)= 0. \label{BC_LE_2D}
\end{equation}
The explicit analytical solution to equation \eqref{LE_2D} is known as
\begin{eqnarray}
	\phi(\rho)&=&2\ln\left[1+\frac{\rho^{2}}{8}\right],\\
	n(\rho)&=&\text{e}^{-\phi(\rho)}=\left(1+\frac{\rho^{2}}{8}\right)^{-2}.
\end{eqnarray}
The total number of masses \emph{per cylinder length} is finite as follows
\begin{equation}
	N_\text{z}=2\pi\int_{0}^{\infty}n(\rho)\, \rho\text{d}\rho=8\pi.
\end{equation}

\subsection{Isothermal sphere model}
Many aspects of the isothermal sphere were studied in detail by \cite{Chandra_1939}. Consider a spherically symmetric system in terms of radius $r$ measured from the center. With this configuration, equation \eqref{g_LE} reduces to the isothermal sphere model
\begin{equation}
	\frac{\text{d}^2\phi}{\text{d}\,r^2}+\frac{1}{2}\frac{\text{d}\phi}{\text{d}r}-\text{e}^{-\phi}=0,
	\label{LE_3D}
\end{equation} 
with BCs
\begin{equation}
	\phi(r=0)=0, \qquad \phi'(r=0)= 0. \label{BC_LE_3D}
\end{equation}
The explicit analytical solution to equation \eqref{LE_3D} is not known.  Near $r=0$, the solution takes the forms  
\begin{eqnarray}
	\phi(r\approx0)&=& -\frac{r^2}{6}+\frac{1}{120}\,r^{4}+\dots,\\
	\rho(r\approx0)&=&1-\frac{r^2}{6}+\dots,
\end{eqnarray}
and, as $r\to\infty$, the asymptotic expressions are
\begin{eqnarray}
    \phi(r\to\infty)&\approx&\ln{\frac{r^{2}}{2}}-\frac{C_1}{r^{1/2}}\cos\left[\frac{\sqrt{7}}{2}\ln{r}+C_2\right],\\
	\rho(r\to\infty)&\approx&{\frac{2}{r^{2}}}\left[1+\frac{C_1}{r^{1/2}}\cos\left(\frac{\sqrt{7}}{2}\ln{r}+C_2\right)\right],
\end{eqnarray}
where $C_{1}$ and $C_{2}$ are constant. The total number of masses is proportional to radius
\begin{equation}
	N=4\pi\int_{0}^{\infty}n(r)\, r^{2}\text{d}r=4\pi r^{2}\phi'(r\to\infty)\propto r.
\end{equation}

\section{Perturbed potentials due to test mass in the isothermal models }\label{Perturb_iso}

We first describe basic ideas of our perturbation method to examine gravitational amplification for the isothermal models, by extending the mathematical formulation used in Section \ref{Deriv_Iso}.  We then derive the Poisson equation for each perturbed isothermal model.

\subsection{Perturbation method for the isothermal systems}\label{Perturb_Isos}

Imagine that test mass $m_\text{p}$ is instantaneously added at the center of the symmetry of one of the isothermal models so that test mass does not break the symmetry of the model. After the perturbation, we may expect the form of DF as follows
\begin{eqnarray}
	F(\boldsymbol{r},\boldsymbol{v})&=&mn_\text{c}\Omega^{D/2}\exp\left[-\frac{E(\boldsymbol{r},\boldsymbol{v})}{k_\text{B}T}\right]+\epsilon f_{1}(\boldsymbol{r},\boldsymbol{v},t),\label{DF}\\
	E(\boldsymbol{r},\boldsymbol{v})&\equiv&E_\text{o}(\boldsymbol{r},\boldsymbol{v})+\epsilon\phi_{1}(\boldsymbol{r},t),
\end{eqnarray}
where $\epsilon$ is the perturbation parameter, and $\epsilon\phi_{1}$ and $\epsilon f_{1}$ are deviations from potential $\Phi_\text{o}$ and DF $mn_\text{c}\Omega^{D/2}\exp\left[-E(\boldsymbol{r},\boldsymbol{v})/k_\text{B}T\right]$. Then, assume that the system reaches a new thermal equilibrium state. The form of $f_{1}$ must be determined so as to satisfy both the high collisionality condition in equation \eqref{C_Bol} and time-independent collisionless Boltzmann equation \eqref{in_col_Bol}. It must be a local Maxwellian DF with constant temperature $T$. The DF $f_{1}$ is described in terms of $E_\text{o}(\boldsymbol{r},\boldsymbol{v})$ at the order of $\epsilon$ and may have an uncertainty by the factor of a constant $A$. In equation \eqref{DF}, expand functions up to the order of $\epsilon$, then
\begin{equation}
	F(\boldsymbol{r},\boldsymbol{v})=F_\text{o}(\boldsymbol{r},\boldsymbol{v})+\epsilon F_\text{o}(\boldsymbol{r},\boldsymbol{v})\left(A-\frac{1}{k_\text{B}T}\phi_{1}(\boldsymbol{r})\right).\label{F_new}
\end{equation}   
In this equation, the contribution of DF $f_{1}$ is only to shift the values of potential $\phi_{1}$ by the constant $Ak_\text{B}T$. Hence, $f_{1}$ is not essential to discuss gravitational fields and may be set to zero;
\begin{equation}
	f_{1}=0. \label{f_1}
\end{equation} 
This treatment for $f_{1}$ distinctly differs from that for collisionless systems. Without the collisionality condition in equation \eqref{C_Bol}, the general solution to time-independent collisionless Boltzmann equation \eqref{in_col_Bol} is any function of isolating integrals \citep{Binney_2011}. \cite{Gilbert_1970} determined the form of $f_{1}$ assuming that test mass gradually appears and that orbital effects are included in an orbit-averaged form of perturbed potential.

Based on the analysis above, we have the Poisson equation perturbed by test mass $m_\text{p}$
\begin{eqnarray}
	\frac{\partial}{\partial \boldsymbol{r}}\cdot\left(\frac{\partial }{\partial \boldsymbol{r}}[\Phi_\text{o}+\epsilon\phi_{1}]\right)&=&4\pi Gmn_\text{c}\exp\left[-\frac{\Phi_\text{o}(\boldsymbol{r})-\Phi_\text{c}+\epsilon\phi_{1}}{k_\text{B}T}\right]\nonumber\\
	&\,&+4\pi \epsilon Gm_\text{p}\delta(\boldsymbol{r}),\label{Eq.pertur_LE}
\end{eqnarray}  
where $\delta(\boldsymbol{r})$ is the delta function of $\boldsymbol{r}$. At the order of $\epsilon$
\begin{equation}
	\frac{\partial}{\partial \boldsymbol{r}}\cdot\left(\frac{\partial \phi_{1}}{\partial \boldsymbol{r}}\right)=-\frac{1}{L_\text{c}^{2}}\exp\left[-\frac{\Phi_\text{o}(\boldsymbol{r})-\Phi_\text{c}}{k_\text{B}T}\right]\phi_{1}+4\pi Gm_\text{p}\delta(\boldsymbol{r}).\label{Poiss_eq_pert}
\end{equation}  
Using the dimensionless variables $\boldsymbol{\xi}$ and $\phi$ defined in equation \eqref{dimless_var}, we obtain a dimensionless form of equation \eqref{Poiss_eq_pert}
\begin{equation}
		\frac{\partial}{\partial \boldsymbol{\xi}}\cdot\left(\frac{\partial \delta\phi}{\partial \boldsymbol{\xi}}\right)+\exp\left[-\phi\right] \delta\phi=\delta(\boldsymbol{\xi}),\label{main_eq}
\end{equation}
where $\delta \phi$ is the dimensionless form of potential $\phi_{1}$ defined as
\begin{equation}
	\delta \phi\equiv\frac{\phi_{1}}{k_\text{B}T},
\end{equation}
and $\epsilon$ is defined as
\begin{equation}
	\epsilon\equiv\dfrac{m_\text{p}}{mn_\text{c}L_\text{c}^{D}}.\label{Eq.eps}
\end{equation}
Equation \eqref{main_eq} must be solved with proper BCs for each isothermal model.

\subsection{Perturbed isothermal sheet}\label{Pertubed_iso_1d}

Assume that test mass sheet is instantaneously added at $z=0$ on the $xy$-plane in the isothermal sheet model, then equation \eqref{main_eq} reduces to
\begin{equation}
	\frac{\text{d}^2\delta\phi}{\text{d}\,z^2}+\text{e}^{-\phi}\delta\phi=\delta(z).
	\label{LE_1D_perturb}
\end{equation} 
We regularize this equation so it becomes numerically solvable. The \emph{raw} potential due to test sheet mass $\delta(z)$ has a singularity at $z=0$ as follows 
\begin{equation}
	\phi_\text{s}(z)=\frac{1}{2}|z|,
\end{equation}
where the reference value of potential constant is set to zero. We remove the singularity by defining a regularized DF
\begin{equation}
	W(z)\equiv\delta\phi(z)-\frac{1}{2}|z|.
\end{equation}
Then, equation \eqref{LE_1D_perturb} reduces to 
\begin{equation}
		\frac{\text{d}^2 W}{\text{d}\,z^2}+\text{e}^{-\phi}\left(W+\frac{1}{2}|z|\right)=0.
		\label{LE_1D_perturb_reg}
\end{equation}
Removing the singularity at $z=0$ means that gravitational field strengths due to $\delta \phi$ and $\phi_\text{s}$ are canceled out near $z=0$. Hence, the BC for potential $W$ is at $z=0$
\begin{equation}
	W'(z=0)=0.\label{BC_1D_x_0}
\end{equation}
Gravitational amplification is expected to only reconfigure a mass distribution, like polarization in plasmas. The effective mass $m^{*}$, the partial sum of masses associated with potential $W(z)$, must be zero as $z\to\infty$. The effective mass is at $z$
\begin{eqnarray}
	m^{*}(z)&=&-2\int_{0}^{z}\text{e}^{-\phi\left(z'\right)}\left(W\left(z'\right)+\frac{1}{2}|z'|\right)\text{d}z',\nonumber\\
	     &=&2\left(W'(z)-W'(0)\right).
\end{eqnarray}
With the BC \eqref{BC_1D_x_0}, we hence obtain a BC at $z\to\infty$
\begin{equation}
	W'(z\to\infty)=0.\label{BC_1D_x_inf}
\end{equation}
We numerically solve equation \eqref{LE_1D_perturb_reg} with the two BCs \eqref{BC_1D_x_0} and \eqref{BC_1D_x_inf}.

\subsection{Perturbed isothermal cylinder}

We repeat a similar analysis of Section \ref{Pertubed_iso_1d} for the isothermal cylinder model. Assume that test linear mass aligned with the $z$-axis is instantaneously added into the isothermal cylinder. The \emph{raw} potential due to the test mass is
\begin{equation}
	\phi_\text{s}(\rho)=\frac{1}{2\pi}\ln \rho,
\end{equation}
where the reference value of potential constant  is set to zero. Define the following regularized DF $W(\rho)$ to remove the singularity of potential $\delta \phi(\rho)$ at $\rho=0$
\begin{equation}
	W(\rho)\equiv\delta\phi(\rho)-\frac{1}{2\pi}\ln \rho.
\end{equation}
With this $W(\rho)$ in the isothermal cylinder, equation \eqref{main_eq} reduces to
\begin{equation}
	\frac{\text{d}^2 W}{\text{d}\,\rho^2}+\frac{\text{d}W}{\text{d}\,\rho}+\text{e}^{-\phi}\left(W+\frac{1}{2\pi}\ln \rho\right)=0.
	\label{LE_2D_perturb_reg}
\end{equation}
The BC for potential $W(\rho)$ is at $\rho=0$
\begin{equation}
	W'(\rho=0)=0.\label{BC_2D_rho_0}
\end{equation}
With this BC, the effective mass is at radius $\rho$ 
\begin{eqnarray}
	m^{*}(\rho)&=&-2\pi\int_{0}^{\rho}\text{e}^{-\phi\left(\rho'\right)}\left(W\left(\rho'\right)+\frac{1}{2\pi}\ln \rho'\right)\text{d}\rho',\nonumber\\
	&=&2\pi\rho W'(\rho).\label{eff_m_cond_2d}
\end{eqnarray}
To make $m^{*}(\rho\to\infty)$ be zero, we need the following BC at $\rho\to\infty$
\begin{equation}
	W'(\rho\to\infty)=0.\label{BC_2D_rho_inf}
\end{equation}
However, $m^{*}(\rho)$ is proportional to $\rho$. Potential $W(\rho\to\infty)$ must asymptotically decay more rapidly than $\rho^{-1}$. The explicit asymptotic expression is readily obtained, by solving equation \eqref{LE_2D_perturb_reg} in the limit of $\rho\to\infty$, as
\begin{equation}
	W'(\rho\to\infty)=\frac{32}{4\pi \rho^{3}}\left(1+3\ln \rho\right).
\end{equation}
With this equation, $m^{*}(\rho\to\infty)$ approaches zero. Hence, equation \eqref{LE_2D_perturb_reg} can be solved with the BCs \eqref{BC_2D_rho_0} and \eqref{BC_2D_rho_inf}.

\subsection{Perturbed isothermal sphere}

Assume that test point mass is instantaneously added at the center of the isothermal sphere. The raw potential due to the mass is
\begin{equation}
	\phi_\text{s}(r)=-\frac{1}{4\pi r}.
\end{equation}
The following regularized DF is defined to remove the singularity of $\delta \phi(r)$ at $r=0$
\begin{equation}
	W(r)\equiv\delta\phi(r)+\frac{1}{4\pi r}.\label{Eq.W_r}
\end{equation}
With this $W(r)$ in the isothermal sphere, equation \eqref{main_eq} reduces to
\begin{equation}
	\frac{\text{d}^2 W}{\text{d}\,r^2}+\frac{1}{2}\frac{\text{d}W}{\text{d}r}+\text{e}^{-\phi}\left(W-\frac{1}{4\pi r}\right)=0.
	\label{LE_3D_perturb_reg}
\end{equation}
Because of the divergent factor, $1/4\pi r$, in equation \eqref{LE_3D_perturb_reg} at $r=0$, a BC for a regular potential $W(r)$ must be taken as
\begin{equation}
	W'(r=0)=\frac{1}{8\pi}.\label{BC_3D_r_0}
\end{equation}
Using this BC, the effective mass is at radius $r$
\begin{eqnarray}
	m^{*}(r)&=&-4\pi\int_{0}^{r}\text{e}^{-\phi\left(r'\right)}\left(W\left(r'\right)-\frac{1}{4\pi r'}\right)\text{d}r',\nonumber\\
	&=&4\pi r^{2} W'(r).\label{eff_m_cond_3d}
\end{eqnarray}
Hence, the following BC at $r\to\infty$ must be satisfied
\begin{equation}
	W'(r\to\infty)=0.\label{BC_3D_r_inf}
\end{equation}
The asymptotic behavior of $W(r)$, however, can not make $m^{*}(r)$ zero as $r\to\infty$. Solving equation \eqref{LE_3D_perturb_reg} as $r\to\infty$ provides the following asymptotic expression
\begin{equation}
	W(r\to\infty)=\frac{C_3}{r^{1/2}}\cos\left[\frac{\sqrt{7}}{2}\ln{r}+C_4\right],
\end{equation}
where $C_3$ and $C_4$ are constant. Accordingly, the effective mass diverges with $r$, like $\sim\sqrt{r}$. This divergence mathematically originates from the slow decay in the isothermal sphere density at large radii $(n\sim2/r^2)$. It can be easily confirmed, by assuming that $W(r)$ decays in a power-law fashion, that the density $n(r)$ must decay more rapidly than $r^{-2}$ to avoid the divergent effective mass. Even though $m^{*}(r)$ diverges, a physically proper BC at $r\to\infty$ is still equation \eqref{BC_3D_r_inf} since it means that  gravitational fields due reconfigured mass distribution vanish at large radii. We solve equation \eqref{LE_3D_perturb_reg} with the BCs \eqref{BC_3D_r_0} and \eqref{BC_3D_r_inf} after obtaining potential $\phi(r)$ by solving equation \eqref{LE_3D} with the BCs in equation \eqref{BC_LE_3D}.

\section{Numerical results}\label{sec:result}
The present Section provides the numerical results for the perturbed isothermal models explained in Section \ref{Perturb_iso}. We introduce the following quantity as a measure of gravitational amplification
\begin{equation}
	\mu=\left|\frac{\delta \phi'(\xi)}{\phi_\text{s}'(\xi)}\right|.
\end{equation}
With this measure, if a perturbed gravitational field, $-\delta \phi'$, is amplified then $\mu>1$ while if it equals the field due to test mass then $\mu=1$. It turns out that the characteristics of gravitational amplification is quite similar between the isothermal sheet and cylinder. We hence first explain their results, and then the result for the isothermal sphere.

\subsection{Numerical results for the isothermal sheet and cylinder}
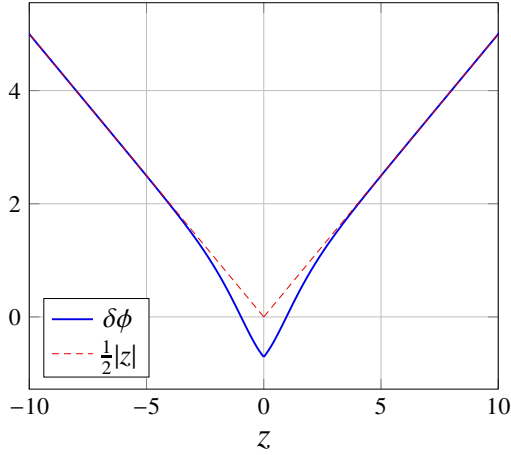
\begin{figure}
	\centering
	\begin{tikzpicture}[scale=0.9]
	\begin{axis}[ grid=major,xlabel=\Large{$z$},xmin=-10, xmax=10, legend pos=south west]
	\addplot [color = bblue ,mark=no,thick,solid ] table[x index=0, y index=2]{Iso1D_r_mr_Phi_Phio.txt}; 
	\addlegendentry{\large{$\delta\phi$}}
	\addplot [color = bred ,mark=no,thin,densely dashed] table[x index=0, y index=3]{Iso1D_r_mr_Phi_Phio.txt}; 
	\addlegendentry{\large{$\frac{1}{2}|z|$}}
	\addplot [color = bblue ,mark=no,thick,solid ] table[x index=1, y index=2]{Iso1D_r_mr_Phi_Phio.txt}; 
	\addplot [color = bred ,mark=no,thin,densely dashed] table[x index=1, y index=3]{Iso1D_r_mr_Phi_Phio.txt}; 
	\end{axis}
	\end{tikzpicture}
\caption{Amplified and raw potentials due to the test mass sheet in the isothermal sheet model.}
\label{fig:1D_phi}
\end{figure}

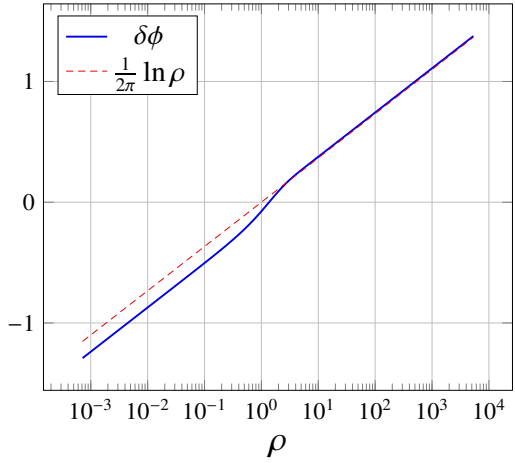
\begin{figure}
	\centering
	\begin{tikzpicture}[scale=0.9]
		\begin{semilogxaxis}[ grid=major,xlabel=\Large{$\rho$},legend pos=north west]
			\addplot [color = bblue ,mark=no,thick,solid ] table[x index=0, y index=1]{Iso2D_r_Phi_Phio.txt}; 
			\addlegendentry{\large{$\delta\phi$}}
			\addplot [color = bred ,mark=no,thin,densely dashed] table[x index=0, y index=2]{Iso2D_r_Phi_Phio.txt}; 
			\addlegendentry{\large{$\frac{1}{2\pi}\ln \rho$}}
		\end{semilogxaxis}
	\end{tikzpicture}
	\caption{Amplified and raw potentials due to the test mass line in the isothermal cylinder model.}
	\label{fig:2D_phi}
\end{figure}

\begin{figure}
	\centering
	\begin{tikzpicture}[scale=0.9]
		\begin{semilogxaxis}[ grid=major,xlabel=\Large{$\rho$}, legend pos=north east]
			\addplot [color = bblue ,mark=no,thick,solid ] table[x index=0, y index=1]{Iso2D_r_Est_Rho.txt}; 
			\addlegendentry{\large{$\mu$}}
			\addplot [color = bred ,mark=no,thin,densely dashed] table[x index=0, y index=2]{Iso2D_r_Est_Rho.txt}; 
			\addlegendentry{\large{$n(\rho)+1$}}
		\end{semilogxaxis}
	\end{tikzpicture}
	\caption{Ratio $\mu$ of the amplified potential to the raw potential due to test line mass in the isothermal cylinder. The density of the isothermal cylinder is plotted as a guide for the far-field decay in the ratio.}
	\label{fig:2D_Est}
\end{figure}
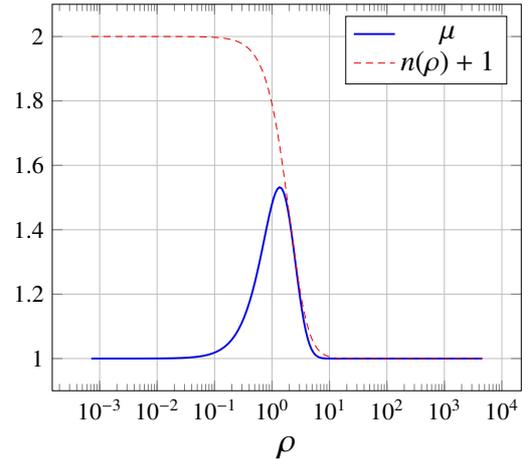

Figures \ref{fig:1D_phi} and \ref{fig:2D_phi} show perturbed potential $\delta \phi(\xi)$ and unperturbed potential $\phi_\text{s}(\xi)$ in the isothermal sheet and cylinder. Figures \ref{fig:1D_Est} and \ref{fig:2D_Est} depict quantity $\mu$ with shifted isothermal sheet- and cylinder- densities. The characteristics of $\mu$ is quite similar to the effective mass  due to test point mass in the isochrone reported in \citep{Gilbert_1970}. The amplification effect is maximized only once within the system and disappears at $\xi=0$ and $\xi\to\infty$. The maximum value of $\mu$ is 1.68 at $z\simeq0.97$ in the isothermal sheet while 1.53 at $\rho\simeq1.4$ in the isothermal cylinder. The potential $\delta\phi$ behaves like the raw potential $\phi_\text{s}$ at large $\xi$. This can be easily understood in equations \eqref{LE_1D_perturb_reg} and \eqref{LE_2D_perturb_reg} in which density dependence becomes weak at large $\xi$.

\begin{figure}
	\centering
	\begin{tikzpicture}[scale=0.9]
		\begin{semilogxaxis}[ grid=major,xlabel=\Large{$z$}, legend pos=north east]
			\addplot [color = bblue ,mark=no,thick,solid ] table[x index=0, y index=1]{Iso1D_r_Est_Rho.txt}; 
			\addlegendentry{\large{$\mu$}}
			\addplot [color = bred ,mark=no,thin,densely dashed] table[x index=0, y index=2]{Iso1D_r_Est_Rho.txt}; 
			\addlegendentry{\large{$n(z)+1$}}
		\end{semilogxaxis}
	\end{tikzpicture}
	\caption{Ratio $\mu$ of the amplified potential to the raw potential due to test sheet mass in the isothermal sheet. The density of the isothermal sheet is plotted as a guide for the far-field decay in the ratio.}
	\label{fig:1D_Est}
\end{figure}
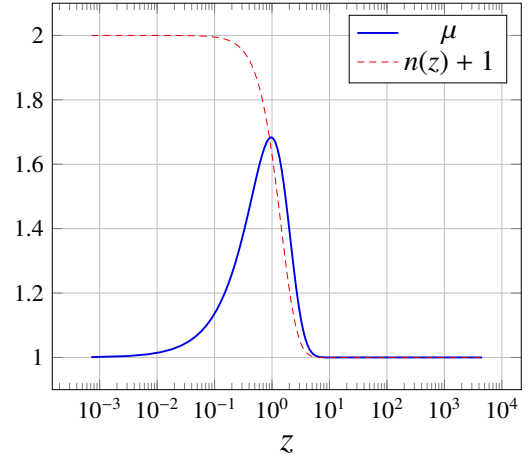

\subsection{Numerical result for the isothermal sphere}

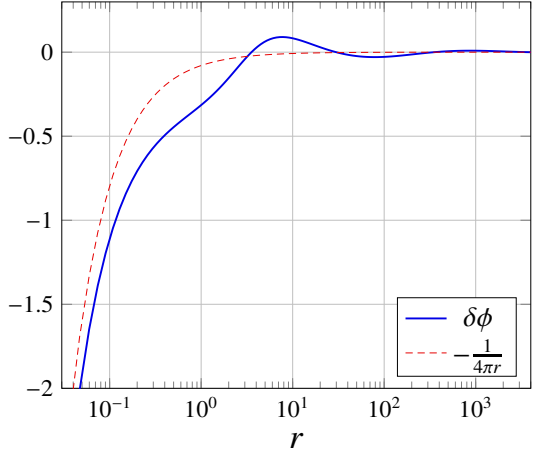
\begin{figure}
	\centering
	\begin{tikzpicture}[scale=0.9]
		\begin{semilogxaxis}[ grid=major,xlabel=\Large{$r$},ymin=-2,xmin=3e-2,xmax=4e3, legend pos=south east]
			\addplot [color = bblue ,mark=no,thick,solid ] table[x index=0, y index=1]{Iso3D_r_Phi_Phio.txt}; 
			\addlegendentry{\large{$\delta\phi$}}
			\addplot [color = bred ,mark=no,thin,densely dashed] table[x index=0, y index=2]{Iso3D_r_Phi_Phio.txt}; 
			\addlegendentry{\large{$-\frac{1}{4\pi r}$}}
		\end{semilogxaxis}
	\end{tikzpicture}
	\caption{Amplified and raw potentials due to the test point mass in the isothermal sphere.}
	\label{fig:3D_phi}
\end{figure}

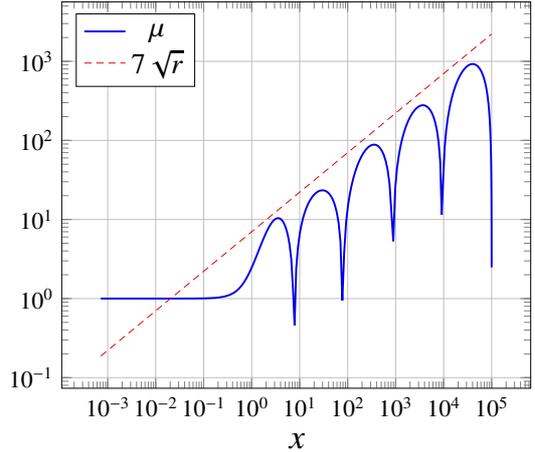
\begin{figure}
	\centering
	\begin{tikzpicture}[scale=0.9]
		\begin{loglogaxis}[ grid=major,xlabel=\Large{$x$}, legend pos=north west]
			\addplot [color = bblue ,mark=no,thick,solid ] table[x index=0, y index=1]{Iso3D_r_Est_Rho.txt}; 
			\addlegendentry{\large{$ \mu $}}
			\addplot [color = bred ,mark=no,thin,densely dashed] table[x index=0, y index=2]{Iso3D_r_Est_Rho.txt}; 
			\addlegendentry{\large{$7\sqrt{r}$}}
		\end{loglogaxis}
	\end{tikzpicture}
	\caption{Ratio of the amplified potential to the raw potential due to test point mass in the isothermal sphere. The asymptotic expression of the isothermal sphere is plotted as a guide for the far-field growth in the ratio.}
	\label{fig:3D_Est}
\end{figure}

Figure \ref{fig:3D_phi} shows the potentials $\delta \phi(r)$ and $\phi_\text{s}(r)$ in the isothermal sphere. A noteworthy result is that there is a maximum of $\delta\phi(r)$ at $r\simeq7.7$. Accordingly, the gravitational field vanishes, or is \emph{shielded}. A possibility of gravitational shielding was suggested in an early work \citep{Saslaw_1968}. Saslaw derived a dumped sinusoidal density by using hydrodynamic equations for the isothermal sphere and assuming a growing perturbation like $\sim e^{\sigma t}$ where $\sigma>0$. Yet, it is not clear whether we should consider \cite{Saslaw_1968}'s and our results as gravitational shielding. Unlike plasma shielding, the perturbed gravitational potentials keep oscillating with radius and do not rapidly converge to a constant value. A similar oscillatory behavior was reported in the singular collisionless isothermal sphere \citep{Murali_1998}. It appears in common that a point-mass perturbation causes an oscillatory perturbed potential in the isothermal spheres. Figure \ref{fig:3D_Est} depicts the quantity $\mu$ and a rescaled asymptotic behavior of the effective mass $m^{*}$. The quantity $\mu$ oscillates with radius, and its amplitude becomes large obeying $(\sim\sqrt{r})$.  Hence, the gravitational amplification becomes indefinitely large with radius. 

It is noteworthy to point out that particles may feel a repulsive force from test mass at some large radii beyond $r\simeq7.7$, considering recent interesting observational results \citep{Bialy_2021}. They reported shell structures moving toward outward around molecular clouds. The onset of such structures can be readily implied from the structure of perturbed potential $\delta \phi(r)$ in the isothermal sphere. However, the present result is not directly applicable to realistic systems. This is because actual dense clouds are formed in a complicated setting (turbulent multi-component non-spherical open system under the effect of  magnetic fields and radiation).  More importantly, the isothermal sphere is not stable with infinite radius. It may be stable against radial perturbation due to pressure confinement \citep{Bonnor_1956}. The isothermal sphere in pressure equilibrium with an ambient medium is stable only at radii $r\lesssim6.5$.  Hence, the repulsive effect may not be seen in such a system. On the one hand, if we consider the isothermal sphere confined by a thermally conductive or insulated wall, it can be stable at $r\lesssim8.99$ or $r\lesssim 34.4$ \citep{Lynden_Bell_1968}. In the next section, we discuss the former case since it can be readily discussed by using the present method.

\section{Discussion}\label{sec:discussion}

The present section examines the relation of the oscillatory gravitational amplification in the isothermal sphere with gravitational instability. To set up a similar situation to \citep{Lynden_Bell_1968}, we confine the isothermal sphere in a spherical wall centered at the center of the isothermal sphere. We then randomly pick up a total amount of mass $m_\text{p}$ from the finite isothermal sphere and add test point mass $m_\text{p}$ at the center. We also assume that the temperature of the finite isothermal sphere is held constant through test-mass perturbation. For numerical calculations, we essentially need to repeat the same numerical calculation explained in Section \ref{Perturb_iso} but with two modifications. First, the BC for $W(z)$ in \eqref{BC_3D_r_inf} must be assigned at the wall radius $r_\text{w}$ in place of $r\to\infty$;
\begin{equation}
	W'(r_\text{w})=0.\label{BC_3D_r_wall}
\end{equation}  
Second, to hold the total mass of the finite isothermal sphere, the definition of potential $W(r)$ in equation \eqref{Eq.W_r} must be redefined by \citep[See e.g.,][]{Gilbert_1970}
\begin{equation}
	W(r)\equiv\delta\phi(r)+\frac{1}{4\pi r}-\frac{\phi_\text{o}(r)}{N(r_\text{w})},\label{Eq.W_r_new}
\end{equation}
where equation \eqref{Eq.eps} is employed to find the relation between $\epsilon$ and $N(r_\text{w})$, the total number of particles within the confined isothermal sphere $(=\frac{4\pi}{m}\int_{0}^{r_\text{w}}r^{2}\rho_\text{o}(r')\text{d}r')$. Then, the perturbed Lane-Emden equation \eqref{LE_3D_perturb_reg} reduces to 
\begin{equation}
	\frac{\text{d}^2 W}{\text{d}\,r^2}+\frac{1}{2}\frac{\text{d}W}{\text{d}r}+\text{e}^{-\phi}\left(W-\frac{1}{4\pi r}-\frac{\phi_\text{o}(r)}{N(r_\text{w})}\right)=0.
	\label{LE_3D_perturb_reg_wall}
\end{equation}
With the above modifications, the total number of particles, temperature, and volume are conserved through test-mass perturbation.
 
We obtained numerical results by solving equation \eqref{LE_3D_perturb_reg_wall} with BCs \eqref{BC_3D_r_0} and \eqref{BC_3D_r_wall}. The gravitational amplification is positive at all radii if the wall radius  is less and equal to 8.9931 (Figure \ref{fig:3D_Est_wall}).  As $r_\text{w}$ approaches 8.9931, the maximum value of amplification is significantly increases near the wall. On the one hand, at $r_\text{w}\geq8.9932$, gravitational fields due to test mass behaves oscillatorily with amplification and repulsive effect (Figure \ref{fig:3D_Est_wall_unst}).

The above results provide a new relation between gravitational amplification and instability. It has been well known that the finite isothermal sphere is unstable at $r_\text{w}\geq8.99$ if it is in contact with thermal bath \citep{Lynden_Bell_1968,Padmanabhan_1990,Katz_2003,Chavanis_2002_2}. Typically the instability is understood based on negative specific heat. The specific heat of the isothermal sphere is positive at $r_\text{w}\leq8.99$ while negative at $r_\text{w}>8.99$. Negative specific heat, however, does not allow the sphere to be in a thermal equilibrium with a thermal bath since it must be always positive in the canonical ensemble. Figures \ref{fig:3D_Est_wall} and \ref{fig:3D_Est_wall_unst} show that gravitational amplification tends to diverge near the wall when $r_\text{w}$ approaches the critical radius $(\approx 8.99331)$. Such a divergence is physically unrealistic and requires nonlinear analysis at least while it may be considered a sign of global isothermal collapse. Our important conclusion is that gravitational amplification occurs at all radii except at the center and surface in a stable finite isothermal sphere in contact with thermal bath while the amplification shows oscillatory structures if the sphere is unstable. Hence, the repulsive effect of gravitational polarization is not realistic in the present configuration. We still need to examine the stability problem of the perturbed isothermal sphere. Some relations have been discussed between gravitational instabilities and density perturbations in finite isothermal spheres based on variational principles \citep[e.g.,][]{Antonov_1985,Padmanabhan_1990}. A detailed discussion of the relations for the present configuration will be discussed in the following paper.  

\begin{figure}
	\centering
	\begin{tikzpicture}[scale=0.9]
		\begin{loglogaxis}[ grid=major, xmax=3e1, xmin=3e-4, xlabel=\Large{$r$},ylabel=\Large{$\mu$}, legend pos=north west]
			\addplot [color = red ,mark=no,thick,solid ] table[x index=0, y index=1]{Iso3D_r_Est_rm4_fixedN.txt}; 
			\addlegendentry{\large{$r_\text{w}=4$}}
			\addplot [color = orange ,mark=no,thick,densely dashed ] table[x index=0, y index=1]{Iso3D_r_Est_rm7_fixedN.txt}; 
			\addlegendentry{\large{$r_\text{w}=7$}}
			\addplot [color = purple ,mark=no,thick,densely dotted ] table[x index=0, y index=1]{Iso3D_r_Est_rm8_9_fixedN.txt}; 
			\addlegendentry{\large{$r_\text{w}=8.9$}}
			\addplot [color = blue ,mark=no,thick,dashed ] table[x index=0, y index=1]{Iso3D_r_Est_rm8_99_fixedN.txt}; 
			\addlegendentry{\large{$r_\text{w}=8.99$}}
			\addplot [color = black ,mark=no,thick,dotted ] table[x index=0, y index=1]{Iso3D_r_Est_rm8_9931_fixedN.txt};
			\addlegendentry{\large{$r_\text{w}=8.9931$}}
		\end{loglogaxis}
	\end{tikzpicture}
	\caption{Ratio of the amplified potential to the raw potential due to test point mass in the finite isothermal sphere of radius $r_\text{w}\leq8.9931$.}
	\label{fig:3D_Est_wall}
\end{figure}

\begin{figure}
	\centering
	\begin{tikzpicture}[scale=0.9]
		\begin{loglogaxis}[ grid=major, xmax=3e2, xmin=3e-4,ymax=2e7,ymin=1e-2, xlabel=\Large{$r$},ylabel=\Large{$\mu$}, legend pos=north west]
			\addplot [color = black ,mark=no,thick,dotted ] table[x index=0, y index=1]{Iso3D_r_Est_rm8_9932_fixedN.txt};
			\addlegendentry{\large{$r_\text{w}=8.9932$}}
			\addplot [color = blue ,mark=no,thick,dashed ] table[x index=0, y index=1]{Iso3D_r_Est_rm9_fixedN.txt}; 
            \addlegendentry{\large{$r_\text{w}=9$}}			
			\addplot [color = purple ,mark=no,thick,densely dotted ] table[x index=0, y index=1]{Iso3D_r_Est_rm15_fixedN.txt}; 
            \addlegendentry{\large{$r_\text{w}=15$}}
			\addplot [color = orange ,mark=no,thick,densely dashed ] table[x index=0, y index=1]{Iso3D_r_Est_rm1e2_fixedN.txt}; 
			\addlegendentry{\large{$r_\text{w}=100$}}
			\addplot [color = red ,mark=no,thick,solid ] table[x index=0, y index=1]{Iso3D_r_Est_rm1e4_fixedN.txt}; 
            \addlegendentry{\large{$r_\text{w}=10^4$}}
		\end{loglogaxis}
	\end{tikzpicture}
	\caption{Ratio of the amplified potential to the raw potential due to test point mass in the finite isothermal sphere of radius $r_\text{w}\geq8.9932$.}
	\label{fig:3D_Est_wall_unst}
\end{figure}

\section{Conclusion}\label{sec:conclusion}

The present work examined gravitational amplification in the equilibrium isothermal sheet, cylinder, and sphere. We assumed a high collision limit of head-on particle collisions. Also, we assumed that the isothermal systems reach new isothermal equilibrium states after test-mass perturbation. Hence, the DFs for the systems are a local Maxwellian DF with constant $T$ before and after the perturbation. Test mass is added to each system in a specific way so that the spatial symmetry holds in the system. We used a kinetic formulation to compare the present work to that for collisionless systems.

The numerical results showed that the gravitational amplification in the isothermal sheet and cylinder are quite similar to a known result for collisionless isochrone. The amplification is zero at $r=0$ and increases with radius until reaching its maximum value. The gravitational field strength is amplified by 68 $\%$ maximally in the isothermal sheet while 53 $\%$ in the isothermal cylinder. At larger radii, the amplification rapidly diminishes obeying the density decay in the systems. 

The isothermal sphere causes an oscillatory gravitational amplification and repulsive effect. However, it is not a consistent result since its effective mass does not reach zero at the infinite radius. Also, the infinite isothermal model is not a stable against radial perturbation. We hence discussed gravitational polarization due to test point mass in the isothermal sphere enclosed by a wall in contact with a thermal bath. The point mass was gathered from the unperturbed finite isothemal sphere so that the total mass is conserved. Our numerical results showed that gravitational fields are amplified at all radii in stable finite isothermal spheres while they are oscillatory showing amplification and repulsive effects in unstable ones. We will analyze the stability of the perturbed finite isothermal sphere based on a variational method. Lastly, a question remains whether the repulsive effect occurs in unstable systems or/and systems with negative specific heats; the question may addressed by discussing a finite isothermal sphere in the microcanonical ensemble which is stable holding a negative specific heat. 

Since the isothermal cylinder and sheet showed an expected feature of gravitational amplification without facing any mathematical difficulty, we may next take off the assumptions used in the present work. For example, we can analyze the nonlinear effect by directly solving equation \eqref{Eq.pertur_LE} without using a linear approximation. We may also examine hydrodynamical effects, accordingly time-dependent systems. With this setting, we can examine conditions that test mass perturbation can drive the equilibrium isothermal sheet and cylinder into new equilibrium states. Also, since it is known that these isothermal systems are stable against pressure confinement \citep{Horedt_1986}, we may analyze the stability of the perturbed systems as well. Lastly, examining the effect of more realistic setting on the gravitational amplification is important, such as multi-component, rotational, magnetic-field, and finite-size effects.



\bibliographystyle{elsarticle-harv} 
\bibliography{science}


\end{document}